\documentclass[preprint,aps,12pt,notitlepage,nofootinbib,tightenlines]{revtex4}
\usepackage{amsmath}
\usepackage{bm}
\usepackage{times}
\usepackage{braket}
\usepackage{color}
\usepackage{epsfig}
\usepackage{slashed}
\usepackage{hyperref}
\usepackage{multirow}
\usepackage{booktabs}
\usepackage{array}
\usepackage{float}
\newcommand{\PB}{\text{B}}

\newcommand{\beq}{\begin{eqnarray}}
\newcommand{\eeq}{\end{eqnarray}}
\newcommand{\be}{\begin{equation}\begin{aligned}}
\newcommand{\ee}{\end{aligned}\end{equation}}

\newcommand{\gev}{\text{GeV}}

\definecolor{Red}{rgb}{1.,0.,0.}

\definecolor{Blue}{rgb}{0.,0.,1.}

\definecolor{nicered}{rgb}{0.7,0.1,0.1}
\definecolor{nicegreen}{rgb}{0.1,0.5,0.1}
\def\lsim{ {\ \lower-1.2pt\vbox{\hbox{\rlap{$<$}\lower6pt\vbox{\hbox{$\sim$}}}}\ } }
\def\gsim{ {\ \lower-1.2pt\vbox{\hbox{\rlap{$>$}\lower6pt\vbox{\hbox{$\sim$}}}}\ } }

\hypersetup{colorlinks,citecolor=nicegreen,linkcolor=nicered}
\begin{document}

\title{Search for single production of vectorlike $B$-quarks at the LHC}
\author{Jin-Zhong Han$^{1}$\footnote{E-mail: hanjinzhong@zknu.edu.cn}, Yao-Bei Liu$^{2}$\footnote{E-mail: liuyaobei@hist.edu.cn}, Lu Xing$^{1}$, Shuai Xu$^{1}$\footnote{E-mail: shuaixu@cqu.edu.cn}}
\affiliation{1. School of Physics and
Telecommunications Engineering, Zhoukou Normal University, Zhoukou 466001, P.R. China\\
2. Henan Institute of Science and Technology, Xinxiang 453003, P.R. China}

\begin{abstract}
New vectorlike quarks have been proposed in many scenarios of  new physics beyond the Standard Model, which address the hierarchy problem and may be potentially discovered at the Large Hadron Collider~(LHC). Based on a model-independent framework,  we propose to search for the vectorlike $B$-quark~(VLQ-$B$) and focus on resonant production via $b$-gluon fusion through chromomagnetic interactions. We then explore the  possible signals of the VLQ-$B$ through  the $B\to tW$ decay mode  at the 14 TeV LHC. After a rapid simulation of signal and background events, the $2\sigma$ excluded regions and the $5\sigma$ discovery reach in the parameter
plane of $\kappa_{B}-M_B$ are obtained at the LHC with an integrated luminosity of 300~(3000) fb$^{-1}$ in the dilepton final states.
\end{abstract}

\maketitle
\newpage
\section{Introduction}
With the running of the Large Hadron Collider (LHC), the exploration of the heavy resonance, such as additional scalars, new gauge bosons and charged fermions, has currently reached the TeV-scale.
Vectorlike quarks (VLQs)  are predicted to provide a dynamical explanation for the large top quark mass  in some popular new physics scenarios beyond the standard model (SM)~\cite{DeSimone:2012fs}, such as little Higgs models~\cite{ArkaniHamed:2002qy}, composite Higgs models~\cite{Agashe:2004rs}, and other new physics theories~\cite{He:1999vp,He:2001fz,Wang:2013jwa,He:2014ora}. These new particles do not receive mass through a Yukawa coupling term  and thus are not excluded by
current searches~\cite{Aguilar-Saavedra:2013qpa}. Based on their electric charges, such VLQs might appear in $SU(2)$ singlets or multiplets, which could lead to rich phenomenology at the current and future high-energy colliders~(see e.g.,~\cite{Atre:2011ae,Buchkremer:2013bha,Barducci:2017xtw,Fuks:2016ftf,Yang:2014usa,Liu:2018hum,Cacciapaglia:2018lld,
Aguilar-Saavedra:2019ghg,Wang:2020ips,Zhang:2017nsn,Han:2017cvu,Liu:2017rjw,Liu:2017sdg,Liu:2019jgp,Tian:2021oey,Moretti:2017qby,Moretti:2016gkr,Carvalho:2018jkq,
King:2020mau,Qin2021,Han:2021kcr,Han:2021lpg,Gong:2019zws,Gong:2020ouh,Choudhury:2021nib}).

Many experimental searches for the VLQs and the constraints on their masses have been conducted.
Current searches by the ATLAS and CMS
collaborations using LHC Run-II data primarily focus on the quantum chromodynamics~(QCD)-induced  pair-production modes of VLQs and lead to lower bounds on the vector-like quark masses of approximately 1-1.5~TeV~\cite{Aaboud:2018wxv, Aaboud:2018xpj,Aaboud:2018saj,Aaboud:2017qpr,Aaboud:2018pii,Sirunyan:2017usq,Sirunyan:2019sza,CMS:2018zkf,CMS:2020ttz,
Sirunyan:2020qvb}, depending on the assumed VLQ decay pattern.
In this paper we focus on the VLQ-$B$ with an electric charge of $-1/3$, which is  the $SU(2)$ singlet and couple exclusively to third-generation SM quarks.
Very recently, the CMS Collaboration presented a search for VLQ-$B$ pair production in the fully hadronic final state~\cite{Sirunyan:2020qvb}, and excluded the  masses up to 1570 and 1450~GeV for 100\% $B\to bh$ and 100\% $B\to tZ$ cases, respectively.

Given the range of these exclusions, this analysis considers a VLQ-$B$ with a mass
greater than 1.3~TeV~\cite{Buckley:2020wzk}.
For such heavy VLQs,  the single production process becomes more important owing to a reduced phase-space and has the added advantage of providing a window to the ultraviolet completion~\cite{Backovic:2015bca,Cacciapaglia:2018qep,Roy:2020fqf,Tian:2021nmj,Deandrea:2021vje}.
A particularly
interesting set of couplings that this may probe are the transition
magnetic moments, whether of electroweak nature or
chromomagnetic. Simple strategies to probe this were developed in
Refs.\cite{Bhattacharya:2007gs,Bhattacharya:2009xg,Nutter:2012an,Belyaev:2021zgq}.
For  transition chromomagnetic
 moments, an excited bottom quark $b^{\ast}$, can be single  produced  at the LHC via
the $b$-gluon fusion process, $bg\to b^{\ast}$~\cite{Nutter:2012an,Belyaev:2021zgq}. Searches for such $b^{\ast}$ quark have been performed at the LHC by the
ATLAS and CMS Collaboration at $\sqrt{s} = 7, ~8$ and 13~TeV~\cite{ATLAS:2013pau,CMS:2015jvf,CMS:2021iuw,CMS:2021mku}.
The future high-luminosity phase of the LHC~(HL-LHC) is expected to extend the sensitivity in searching for the bounds
on the VLQs mass and couplings. In this  paper, we focus on  the VLQ-$B$ being  the $SU(2)$ singlet and coupling exclusively to third-generation SM quarks.
We consider the $s$-channel-resonant production of the VLQ-$B$ at the 14~TeV LHC via the considered chromomagnetic moment, and present a careful simulation of the signals and SM backgrounds for the subprocess $bg\to B\to tW^{-}$ via the dilepton final state.

This remainder of this paper is organized as follows: in section II, we briefly describe the most general Lagrangian, which is related to the couplings of VLQ-$B$ with the SM particles, and we discuss the decay width in narrow scenarios, the branching ratio and its single production at the 14 TeV LHC.
Section III devotes to a detector simulation  of the signal and the relevant backgrounds at the LHC and its high-luminosity
operation phase. We perform a
detailed collider analysis to extract the projected sensisitivity of the HL-LHC in probing the parameter space through direct searches in the $B\to tW$ decay channel. Finally, we summarize our results in Section IV.

\section{Decays and single production of VLQ-$B$}
\subsection{Effective Lagrangian}
The general Lagrangian describing the effective interaction of VLQ-$B$, can be expressed as~\cite{Nutter:2012an}~
\begin{eqnarray}
\mathcal{L}&=&\frac{g_s}{2\Lambda}G_{\mu\nu}\overline{b}\sigma^{\mu\nu}\Big(\kappa^b_L P_L+\kappa^b_R P_R\Big)B+\frac{g_2}{\sqrt{2}}W_{\mu}^{-} \overline{t} \gamma^{\mu}\Big(Y_LP_L+Y_RP_R\Big)B \nonumber\\
&+&\frac{g_2}{2c_W}Z_{\mu}\overline{b}\gamma^{\mu}\Big(F_L P_L+F_R P_R\Big)B+\frac{m_b}{\upsilon}h\overline{b}\Big(y_L P_L+y_R P_R\Big)B+H.c.. \label{Lag1}
\end{eqnarray}
Here $\Lambda$ is the cutoff scale, often set to the VLQ-$B$  mass; $g_s$ is the  strong coupling constant; $G_{\mu\nu}$ is the field strength tensor of the gluon; $g_2$ is the $SU(2)_L$ coupling constant; and $\upsilon= 246$ GeV is the electroweak scale. The factors $\kappa^b_{L,R}$, $Y_{L,R}$, $F_{L,R}$ and $y_{L,R}$ parameterize the chirality of the VLQ-$B$ couplings with the different SM particles. When the singlet  VLQ-$B$ mixes only with the left-handed bottom quark, we obtain  $Y_R~(F_R) \simeq 0$ and $y_R \simeq \frac{M_B}{m_b} y_L$ . For simplicity, we consider a benchmark scenario with couplings $Y_L =F_L = y_{L}=s_L= {\upsilon}/{M_B}$. For our objectives, only the
magnitude of $\kappa^{b}= \kappa^b_{L,R}$ is relevant (for more details see
Ref.~\cite{Nutter:2012an} and the references therein),  and in the absence of any
theoretical knowledge, we consider a phenomenologically
guided limit $|\kappa^b|\leq 0.5$.

Very recently, the CMS collaboration have searched for a heavy resonance decaying into a top quark and
a $W$ boson in the all-hadronic final state~\cite{CMS:2021iuw} and lepton+jets final state~\cite{CMS:2021mku}, respectively. For a benchmark value of the chromomagnetic transition moments~($\kappa^{b} =1$), the hypotheses of $B$ quarks with left-handed, right-handed, and vector-like chiralities are excluded at the 95\% confidence level for masses below 2.6~(3.0), 2.8~(3.0), and 3.1~(3.2) ~TeV, respectively. Certainly, for a smaller value of coupling parameter $\kappa^{b}$, such mass limits should be reduced accordingly.

\subsection{Decays width and branching ratio}
From Eq. (\ref{Lag1}), we observe that the VLQ-$B$ decays
are dominated by two-body final states, namely $B\to bg$, $B\to tW$, $B\to Zb$ and $B\to~hb$. The first of these proceeds through
the chromomagnetic moment $\kappa^b$. The remain three are driven by the mixing parameters, and for a very large $M_B$, they have nearly
equal partial widths, a consequence of the Goldstone equivalence theorem~\cite{ET-hjh,He:1992nga,He:1993yd,He:1994br,He:1996rb,He:1996cm}. The partial decay widths of the VLQ-$B$ can be epressed as
\begin{eqnarray}
	\Gamma (B \to b\, Z) &=&   \frac{g_2^2}{128\pi c_W^2} \frac{M_{B}^3}{M_Z^2} (F_L^2 + F_R^2) (1 - x_Z^2 )^2 ( 1 +  2 x_Z^2 ), ,\\
	\Gamma (B \to t \,W^- ) &=& \frac{g_2^2}{64\pi} \frac{M_{B}^3}{M_W^2} (Y_L^2 + Y_R^2) (1  -  x_t^2)^3 + {\mathcal O}(x_W^2) \, ,\\
	\Gamma (B \to b \,h ) &=& \frac{g^2}{128\pi} \frac{M_{B}^3}{M_W^2} (y_L^2 + y_R^2) (1 - x_h^2)^2 \, ,\\
	\Gamma (B \to b \,g ) &=& \frac{g_s^2}{12\pi} M_{B}  (\kappa_L^2 + \kappa_R^2) \, .
\end{eqnarray}
Here $x_Z = M_Z/M_{B}$,  $x_W = M_W/M_{B}$, $x_h = M_h/M_{B}$ and $x_t = M_t/M_{B}$.

\begin{figure}[htb]
\begin{center}
\vspace{-0.5cm}
\centerline{\epsfxsize=9cm \epsffile{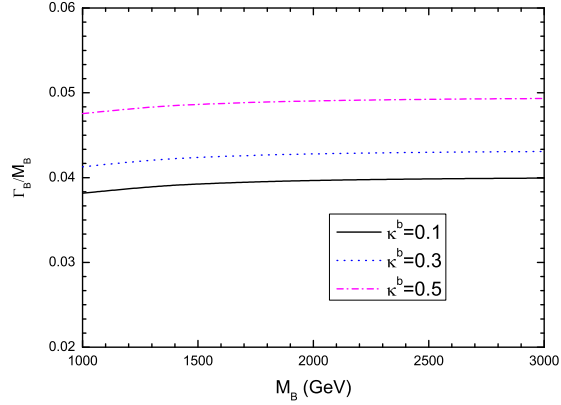}}
\caption{ Width-over-mass ratio $\Gamma_{B}/M_{B}$ as a function of $M_B$ for different coupling strengths $\kappa^b$. }
\label{decay}
\end{center}
\end{figure}

For the  fixed coupling parameters $Y_L =F_L = y_{L}={\upsilon}/{M_B}$, the decay width of the VLQ-$B$ depends on the coupling parameter $\kappa^b$ and its mass $M_B$. In Fig.~\ref{decay}, we depict the dependence of width-over-mass ratio $\Gamma_{B}/M_B$ on $M_B$ for three values of  $\kappa^b$.  We can observe that, even for a very large value of
$\kappa^b$, the ratio $\Gamma_{B}/M_B\lesssim0.05$, thereby validating the
narrow width approximation.
\begin{figure}[htb]
\begin{center}
\vspace{-0.5cm}
\centerline{\epsfxsize=8cm \epsffile{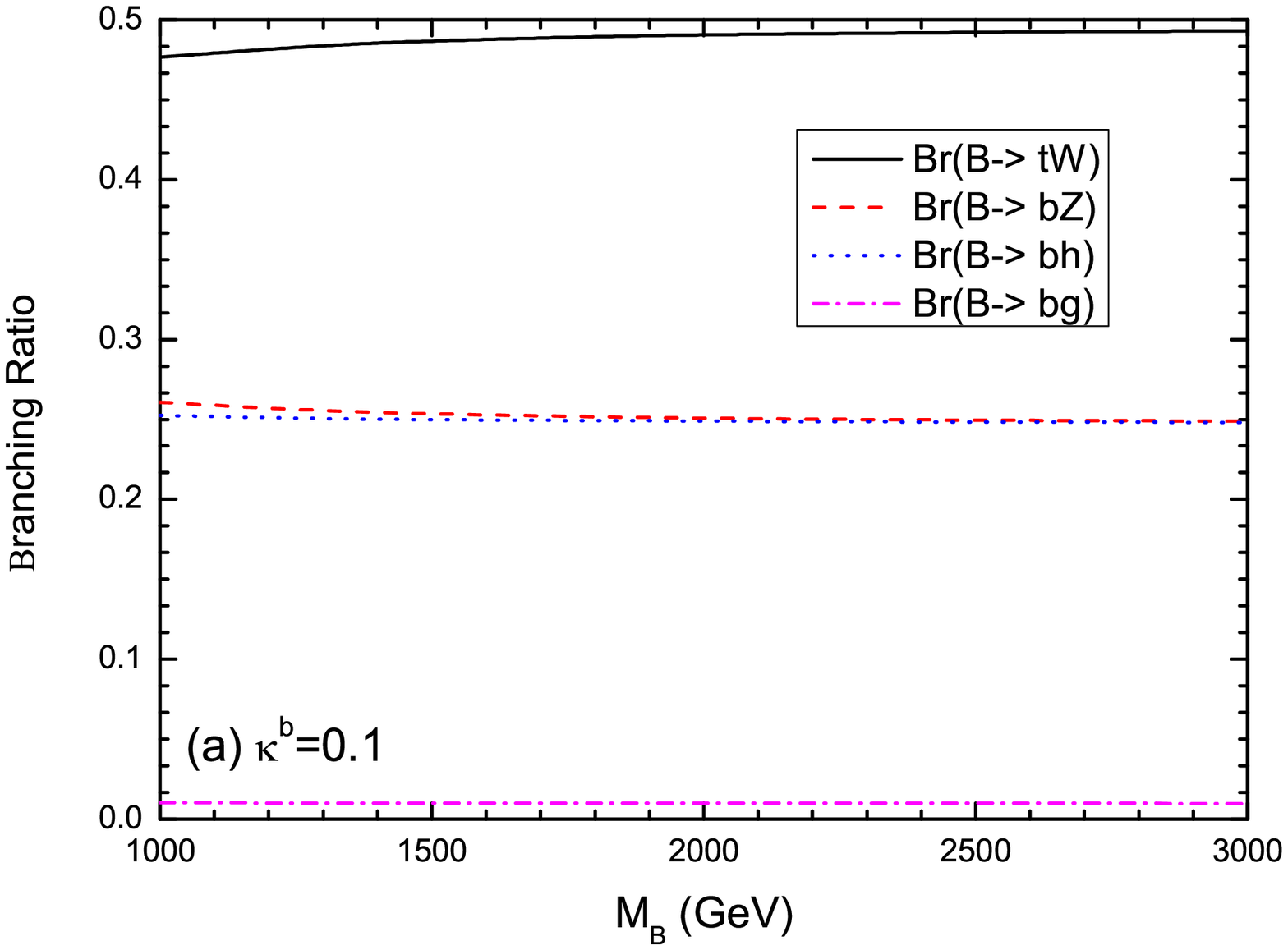}\epsfxsize=8cm \epsffile{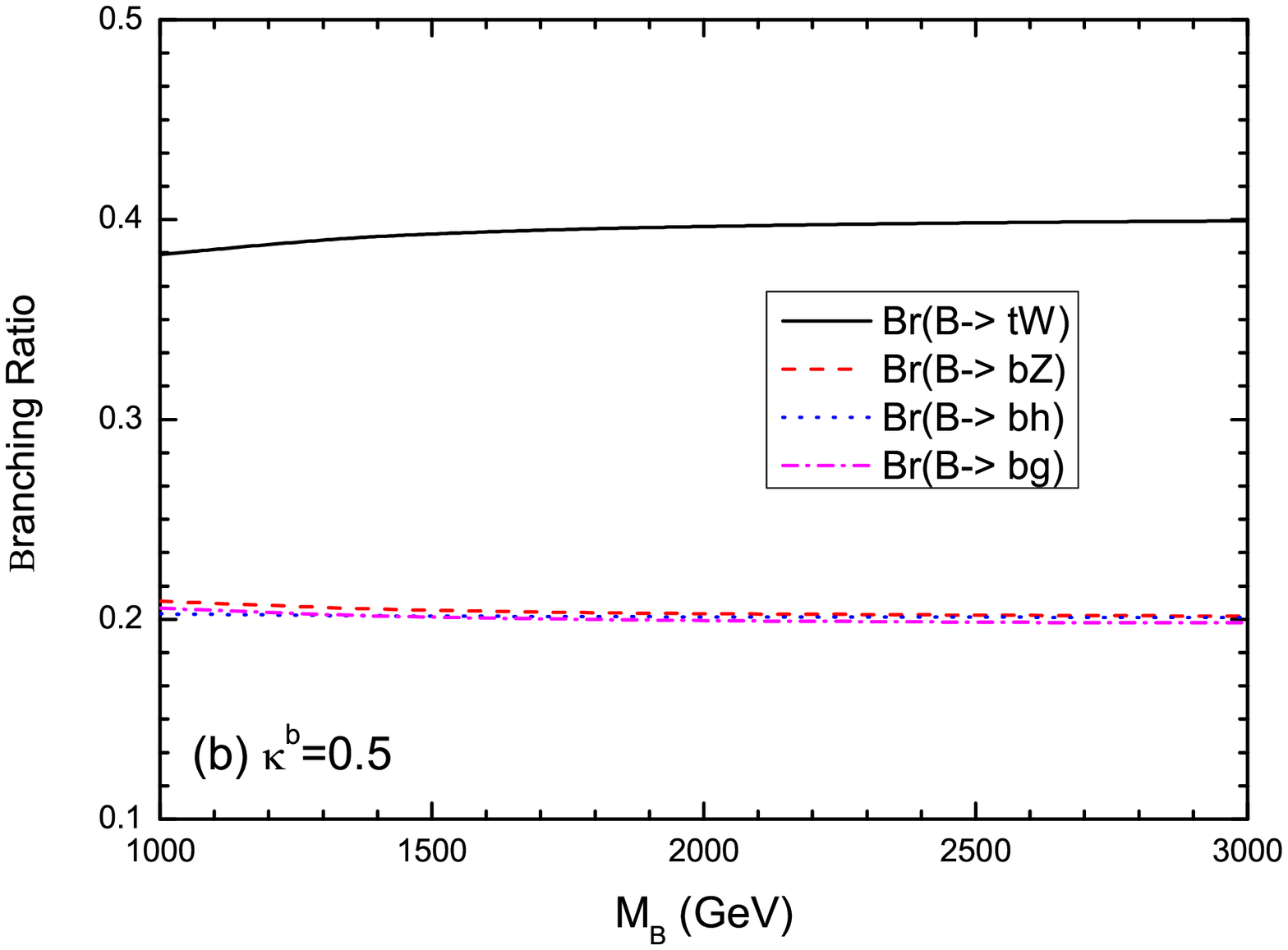}}
\caption{ Branching ratios for the decay modes $Wt$, $Zb$, $hb$ and $bg$ as functions of $M_B$ for  $\kappa^b= 0.1$~(left) and $\kappa^b= 0.5$~(right).}
\label{br}
\end{center}
\end{figure}

The branching ratios for these decay channels  are plotted as functions of the mass parameter $M_B$ in Fig.~\ref{br} for $\kappa^b= 0.1$ and $\kappa^b= 0.5$. We can observe from Fig.~\ref{br} that, for $\kappa^b= 0.1$,  the branching ratio of the decay mode $Wt$ will increase to approximately 50\%, and we obtain $Br(B \rightarrow Zb) \simeq Br(B \rightarrow hb) $. Certainly, if we take a large value $\kappa^b=0.5$, the value of the branching ratio $Br(B \rightarrow bg)$ will be increased to approximately 20\%. This is because for a fixed VLQ-$B$ mass, the partial decay width $\Gamma_{B\to bg}$ is always proportional to $(\kappa^{b})^{2}$.

\subsection{Single production at the LHC}
\begin{figure}[hbtp]
	\centering
    	\includegraphics[width=1.05\textwidth]{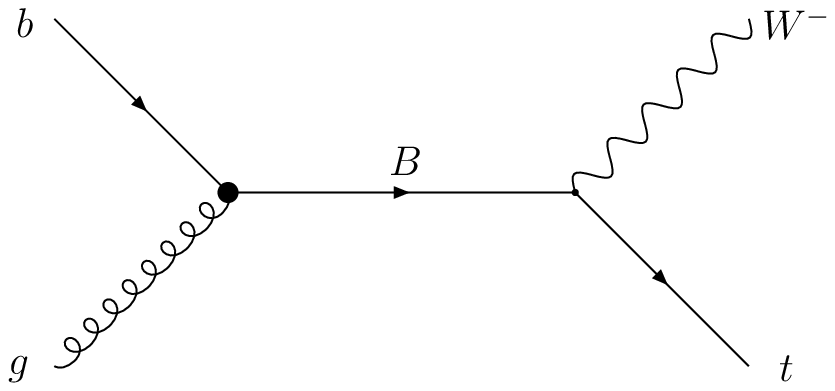}
    \vspace{-19cm}
    	\caption{Feynman diagrams for production of a  vector-like \PB~quark and decay to a $W$ boson and top quark.}
	\label{fey}
\end{figure}
\begin{figure}[h]
\begin{center}
\vspace{-0.5cm}
\centerline{\epsfxsize=10cm \epsffile{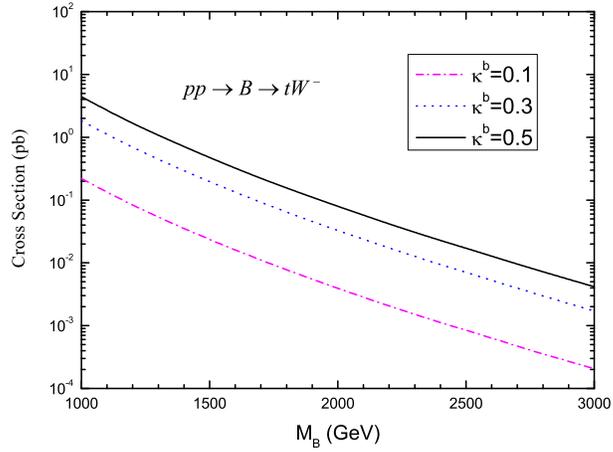}}
\caption{Production cross section of the process $pp\rightarrow B \rightarrow tW$ as a function of  $M_B$ for three values of  $\kappa^b$  at the 14~TeV LHC. }
\label{cross}
\end{center}
\end{figure}
Depending on the strength of the
chromomagnetic interactions $\kappa^b$, the produced VLQ-$B$ can dominantly decay into a $tW$ system, the  Feynman diagram for such resonance VLQ-$B$ production and decay to $tW$ is shown in Fig.~\ref{fey}. The production cross section $ \sigma(pp\rightarrow B \rightarrow tW )$ is plotted in Fig.~\ref{cross}, as a function of the  mass $M_B$ for three typical values of the parameter $\kappa^b$  at the 14 TeV LHC.
The leading-order~(LO) cross sections are obtained by using MadGraph5-aMC$@$NLO~\cite{Alwall:2014hca} with NNPDF23L01 parton distribution functions (PDFs)~\cite{Ball:2014uwa} taking the renormalization and factorization scales of $\mu_R=\mu_F=\mu_0/2=M_B$.
For $0.1\leq\kappa^b \leq0.5$ and 1000 GeV $<M_B<$ 3000 GeV, the values of the cross section  $ \sigma(pp\rightarrow B \rightarrow tW )$ are in the range of $2.1\times10^{-4} \sim 4.4~  $pb. For $\kappa^b=0.3$ and $M_B$ = 1500, 1800, 2000~GeV, their values are 0.2, 0.064, 0.032 $\rm pb$, respectively.
Certainly, increasing $\kappa^b$ would  enhance
the partial decay width $\Gamma(B \to b g)$, thereby suppressing the
branching fraction for others.

\section{Collider simulation and analysis}
Next, we  explore the discovery potentiality of the singlet VLQ-$B$
through the final states in which both the top quark and the $W$ boson originating from the initial VLQ-$B$ decay  leptonically.
\beq\label{signal}
 pp\to B \to t(\to b\ell^{+}\nu)W^{-}(\to \ell^{-}\bar{\nu}),
\eeq
 where $\ell= e, \mu$. Note that the analysis has included the charge-conjugate process.
 As a reference point, we set
a benchmark value of $\kappa = 0.1$. Analogously, our benchmark
points (BP) in the mass-axis read $M_B=$1500, 1800 and 2000~GeV.  However, later, we will present the reach in the $M_B$--$\kappa^b$
plane.

  For the above dilepton final states, the main backgrounds are  top pair production, the associated $tW$
production, and diboson~($WW$, $WZ$ and $ZZ$) production in association with jets~($VV$+jets). Here, we do not consider multijet backgrounds in which jets can fake  electrons as they are generally negligible in multilepton analyses~\cite{Khachatryan:2014ewa}.

We calculate the leading order~(LO) production cross sections and
events of signal and backgrounds at parton level using MadGraph5-aMC$@$NLO. The relevant SM input parameters were obtained from~\cite{pdg}. We then transmitted
the parton-level events to Pythia 8~\cite{pythia8} for showering and
hadronization.
All produced jets were forced to be clustered using  FASTJET 3.2~\cite{Cacciari:2011ma} assuming the anti-$k_{t}$ algorithm with a cone radius of $R=0.4$~\cite{antikt}. Detector effects were simulated with Delphes 3.4.2~\cite{deFavereau:2013fsa}, using the standard HL-LHC detector parameterization shipped with the program.
Finally, event analysis was performed using MadAnalysis5~\cite{ma5}.  To consier inclusive QCD contributions, we generated
the hard scattering of backgrounds with up to
one or two additional jets in the final state, followed by matrix
element and parton shower merging with the MLM
matching scheme~\cite{Frederix:2012ps}. For the SM backgrounds, we generated LO samples renormalized to the NLO or next-NLO (NNLO) order cross sections, which are listed in table~\ref{kf}.

\begin{table}[htb]
\begin{center}
\caption{$K$-factors of the leading SM background processes for our analysis. \label{kf}}
\vspace{0.2cm}
\begin{tabular}{|c|c|c|c|c|}
\hline
Process& $t\bar{t}$ & $tW$ &$VV$+jets& $Z$+jets\\  \hline
$K$-factor&1.6~\cite{Czakon:2013goa}&1.35~\cite{Kidonakis:2016sjf}&1.67~\cite{Campbell:2011bn}&1.2~\cite{Catani:2009sm}\\
\hline
\end{tabular} \end{center}\end{table}

The basic cuts at parton level for the signal and SM backgrounds were taken as follows:
 \be
p_{T}^{\ell}>~20~\gev,\quad
p_{T}^{j}>~25~\gev,\quad
 |\eta_{\ell/j}|<~2.5, \quad
 \Delta R_{ij}>&~0.4 \\
  \ee
where $\Delta R=\sqrt{\Delta\Phi^{2}+\Delta\eta^{2}}$ is the separation in the rapidity-azimuth plane, and $p_{T}^{\ell, j}$ are the transverse momentum of leptons and jets, respectively.

\begin{figure}[htb]
\begin{center}
\centerline{\hspace{2.0cm}\epsfxsize=9cm\epsffile{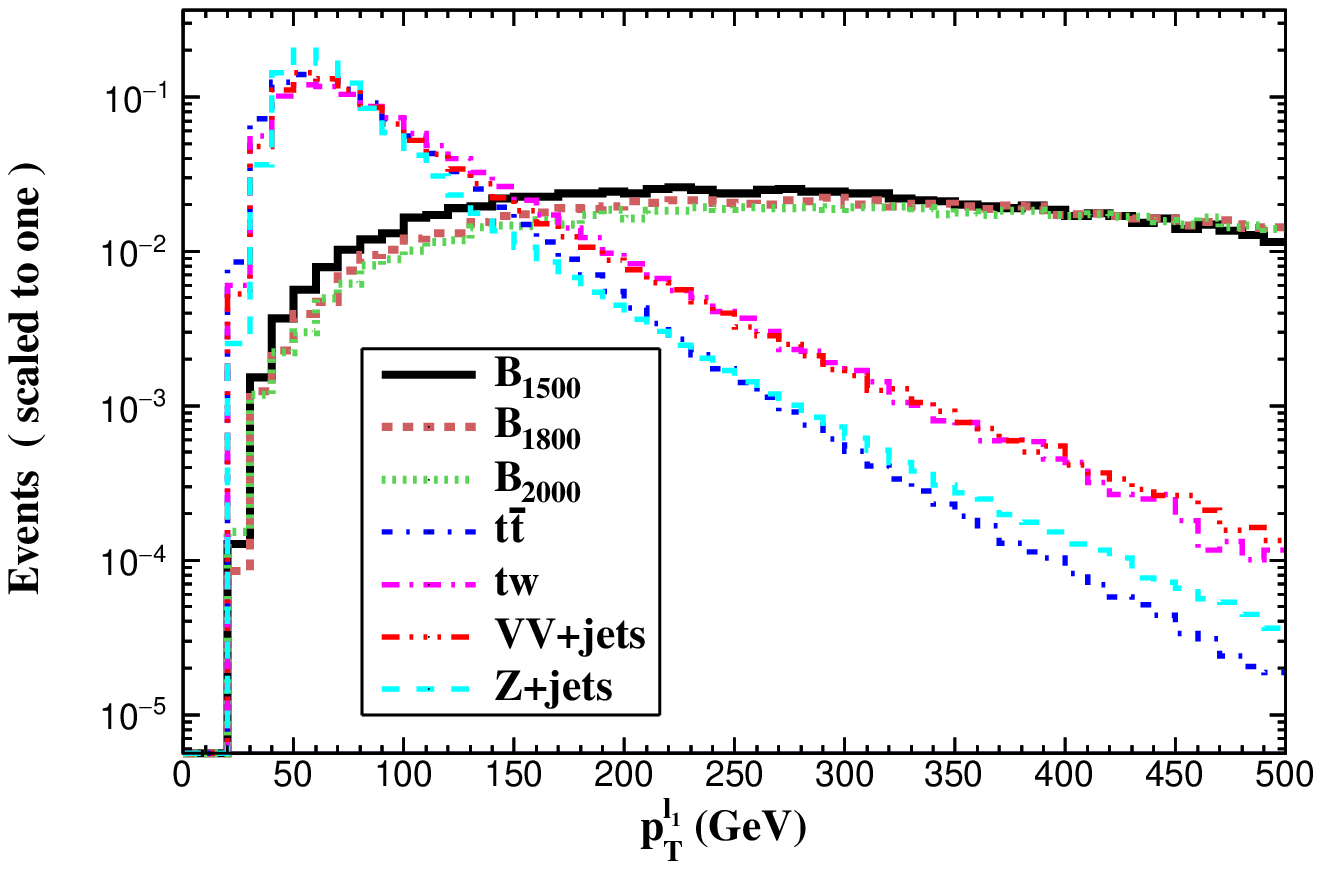}
\hspace{-2.0cm}\epsfxsize=9cm\epsffile{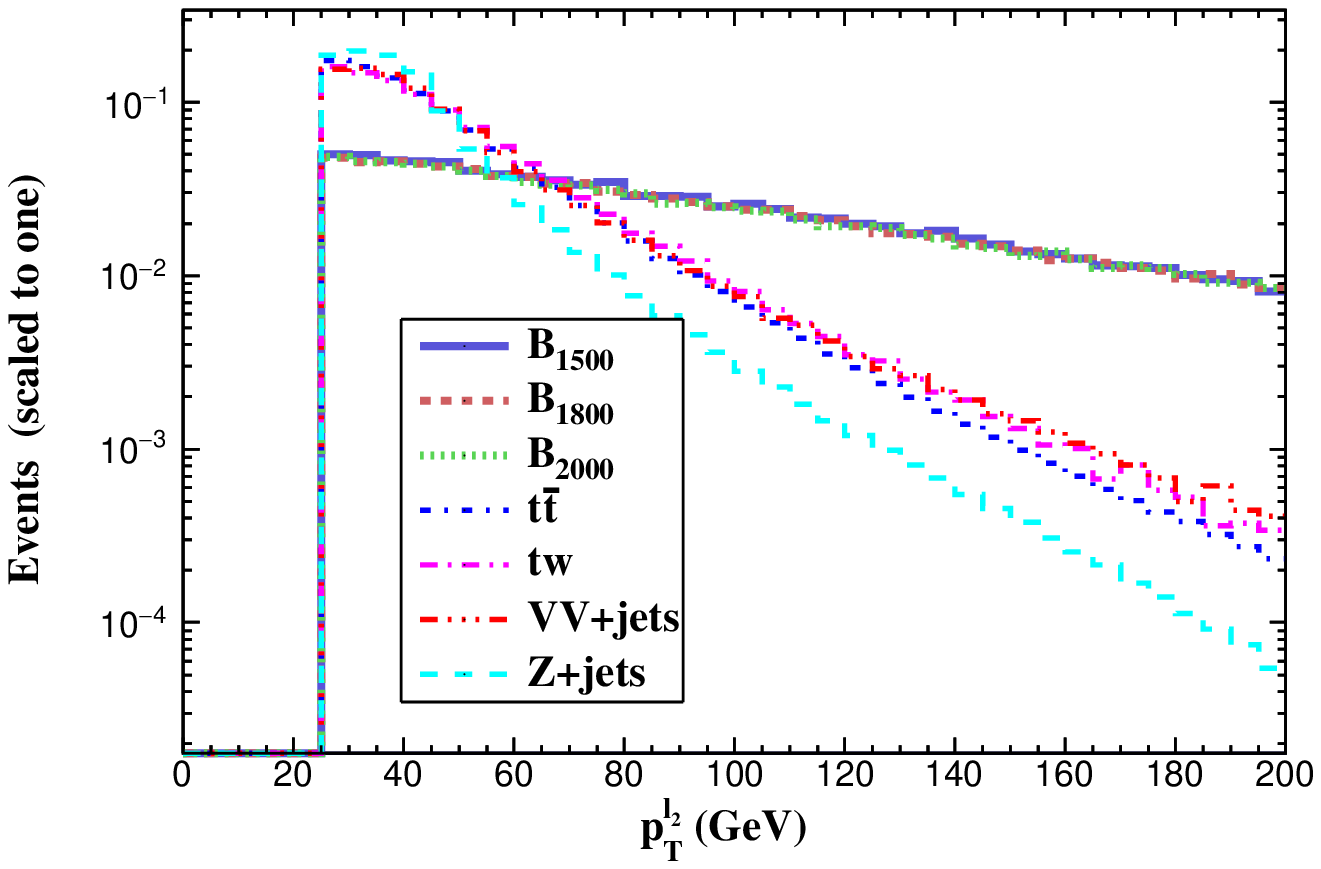}}
\centerline{\hspace{2.0cm}\epsfxsize=9cm\epsffile{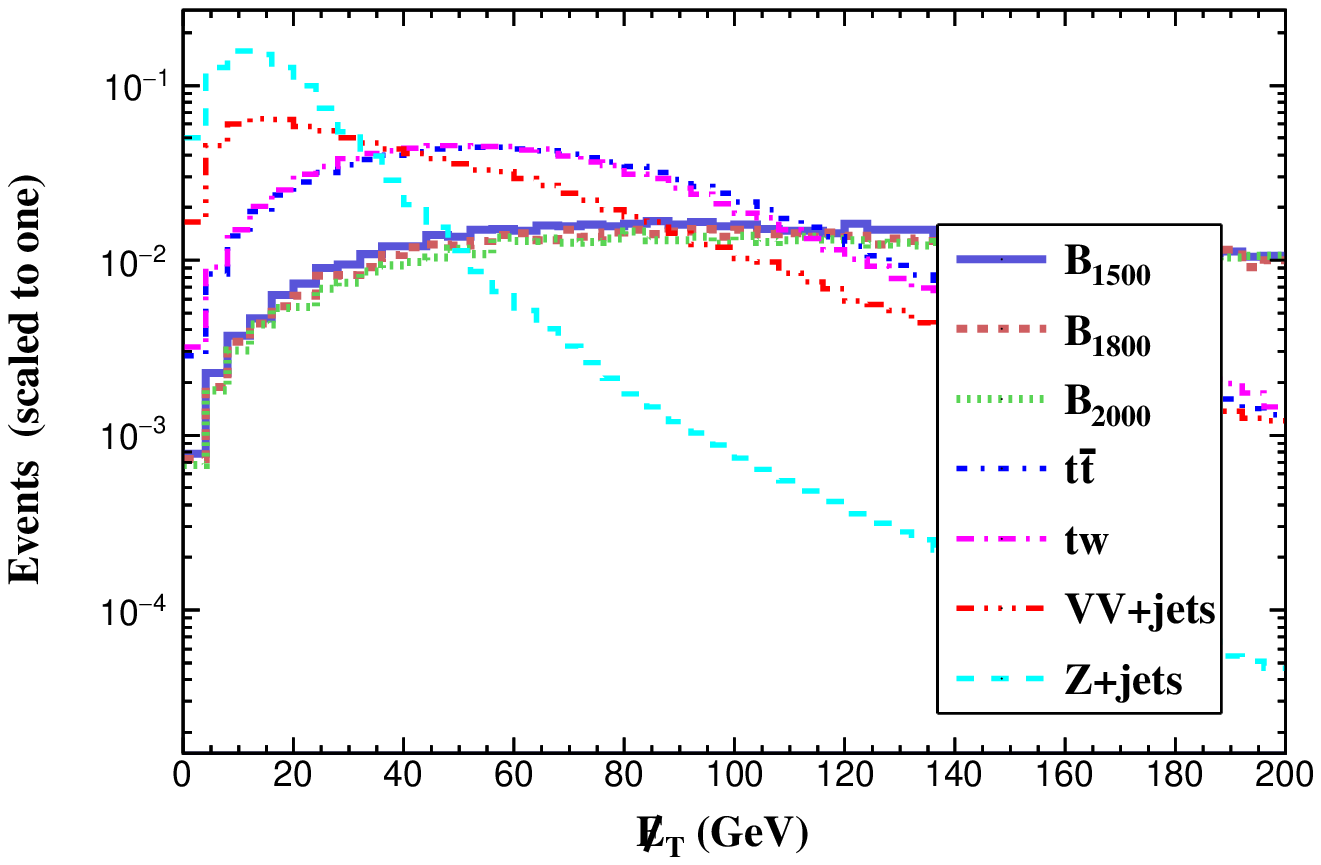}
\hspace{-2.0cm}\epsfxsize=9cm\epsffile{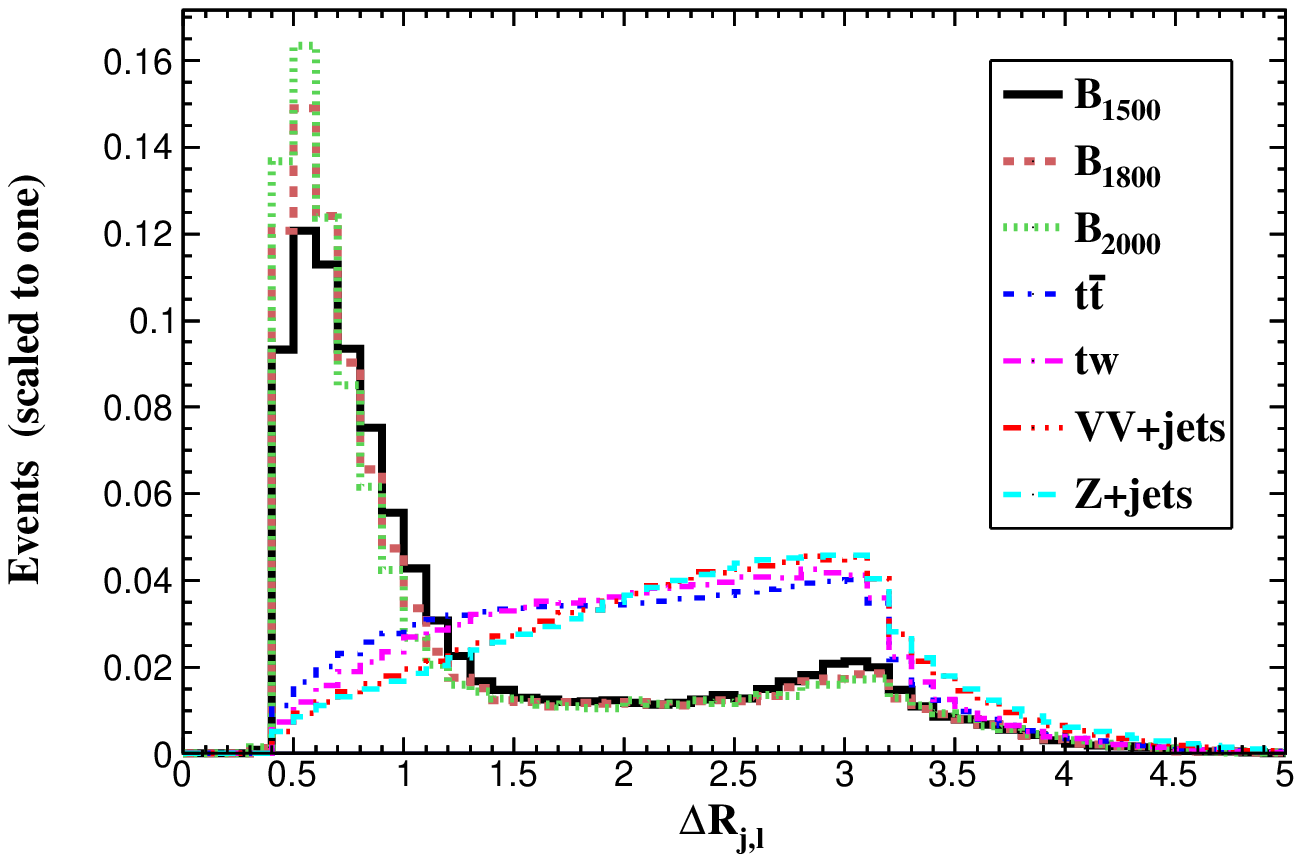}}
\centerline{\hspace{2.0cm}\epsfxsize=9cm\epsffile{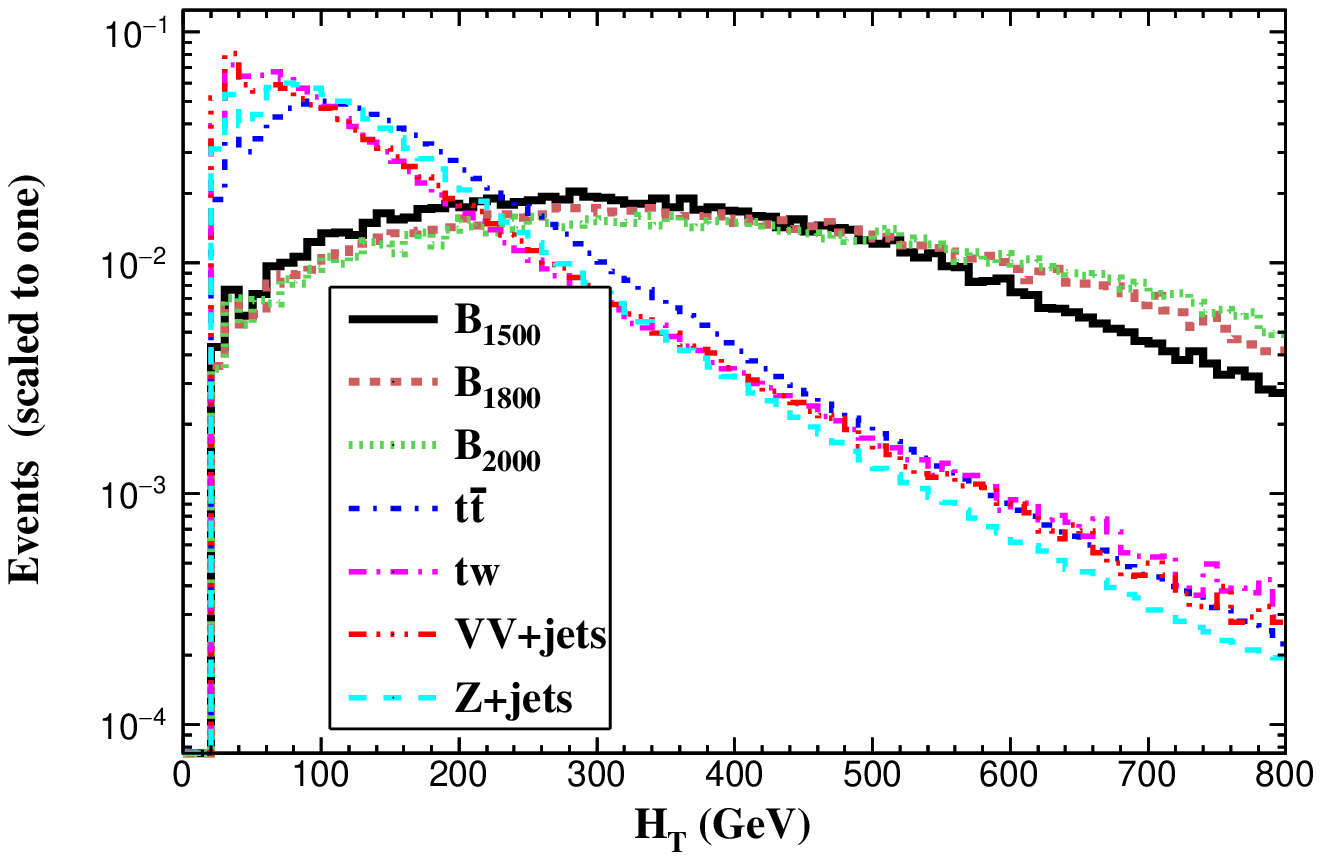}
\hspace{-2.0cm}\epsfxsize=9cm\epsffile{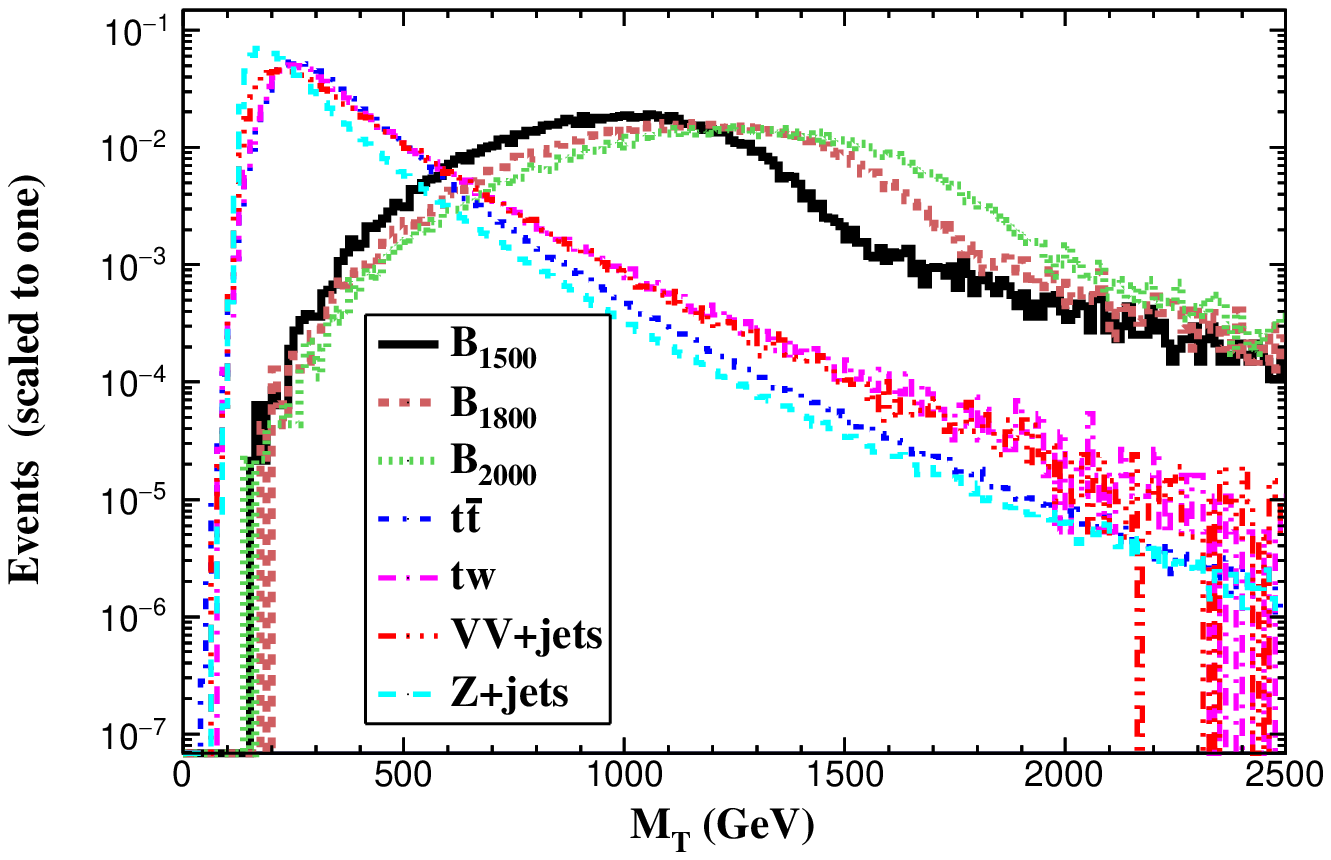}}
\caption{Normalized distributions for the signals (with $M_{B}$=1500, 1800 and 2000~GeV) and SM backgrounds at the 14 TeV LHC. }
\label{distribution1}
\end{center}
\end{figure}

For the signal, the final state is required to have exactly two leptons with opposite
charge and one central jet. No $b$-tagging is require because the dominant background
from $t\bar{t}$ production also contains $b$ quarks.
In Fig.~\ref{distribution1}, we plot some differential distributions  for  signals and SM backgrounds after the basic cuts at the 14 TeV LHC, such as the transverse momentum distributions of the leading and sub-leading leptons~($p_{T}^{\ell_{1},\ell_{2}}$),  missing transverse momentum $\slashed E_T^{miss}$, separations $\Delta R_{j,\ell}$,  scalar sum of the transverse energy of all final-state jets $H_{T}$, and  transverse mass distribution $M_T(j\ell_{1}\ell_{2}\slashed E_T)$. Here, to reconstruct the VLQ-$B$ mass, we use a cluster transverse mass, defined as~\cite{Conte:2014zja}
\beq
M_{T}^{2}(j\ell_{1}\ell_{2}\slashed E_T)=(\sqrt{p_{T}^{2}(j\ell_{1}\ell_{2})+M^{2}_{j\ell_{1}\ell_{2}}}+\slashed E_T)^{2}-(\vec{p}_{T}(j\ell_{1}\ell_{2})+\slashed E_T)^{2},
\eeq
where $\vec{p}_{T}(j\ell_{1}\ell_{2})$ is the total transverse momentum of all  visible particles  and $M_{j\ell_{1}\ell_{2}}$ is their invariant mass. Owing to the larger mass of VLQ-$B$, the decay products of VLQ-$B$ are highly boosted. Therefore, the $p_{T}^{\ell}$ peaks of the signals are larger than those of the SM backgrounds, and the angular
distance between the jet and the lepton is smaller than that in background process events.

Based on these kinematical distributions, we can impose
 the following set of cuts to enhance the signal significance.:
 \begin{itemize}
\item Cut-1: There are exactly two isolated leptons with $p_{T}^{\ell_{1}}> 150 \rm ~GeV$, and $p_{T}^{\ell_{2}}> 60 \rm ~GeV$, and at least one jet is required $p_T> 30$~GeV.
\item Cut-2:
     The missing transverse momentum $\slashed E_T^{miss}$ is required to be larger than 60~GeV.
\item Cut-3:
     The the separations $\Delta R_{j,\ell}$ and $\Delta R_{\ell_{1},\ell_{2}}$ are required to have $\Delta R_{j,\ell}< 1.0$ and $\Delta R_{\ell_{1},\ell_{2}}> 2.5$.
\item
Cut-4: The scalar sum of the transverse energy of all final-state jets is required $H_T>300 \rm ~GeV$.
\item
Cut-5: The transverse mass is required $M_{T}>1000 \rm ~GeV$.
\end{itemize}

\begin{table}[ht!]
\fontsize{12pt}{8pt}\selectfont \caption{Cut flow of the cross sections (in fb) for the signals and SM backgrounds at the 14 TeV LHC with $\kappa^{b}=0.1$ and three typical $B$ quark masses. \label{cutflow}}
\begin{center}
\newcolumntype{C}[1]{>{\centering\let\newline\\\arraybackslash\hspace{0pt}}m{#1}}
{\renewcommand{\arraystretch}{1.5}
\begin{tabular}{C{2.0cm}| C{2.0cm} |C{2.0cm} |C{2.0cm}|C{1.6cm}C{1.6cm}C{1.6cm}C{1.6cm} }
\hline
 \multirow{2}{*}{Cuts}& \multicolumn{3}{c|}{Signals}&\multicolumn{3}{c}{Backgrounds} \\ \cline{2-8}
&1500~GeV &1800~GeV& 2000~GeV  & $t\bar{t}$ & $tW$&$VV$+jets &$Z$+jets  \\   \cline{1-8} \hline
Basic&0.57&0.17&0.077&11700&990&940&41760\\
Cut-1&0.35&0.11&0.05&520&68&54&1012\\
Cut-2 &0.31&0.094&0.044&302&39&24&136\\
Cut-3 &0.17&0.056&0.026&82&7.8&2.2&0.005\\
Cut-4 &0.10&0.036&0.017&26&1.75&0.54&0.002\\
Cut-5 &0.051&0.024&0.013&1.17&0.26&0.29&0.0002
\\
\hline
\end{tabular} }
 \end{center}
 \end{table}

We present the cross sections of three typical signal ($M_B=1500, 1800, 2000$~GeV) and the relevant
backgrounds after imposing
the cuts in Table~\ref{cutflow}. We can observe that, at the end of the cut flow, the backgrounds are suppressed very efficiently, and the largest SM  background is the $t\bar{t}$ process with the cross section of approximately 1.17~fb.

To analyze the observability, we use the median significance  to estimate the expected discovery and exclusion significance~\cite{Cowan:2010js}:
\be
\mathcal{Z}_\text{disc} &=
  \sqrt{2\left[(s+b)\ln\left(\frac{(s+b)(1+\delta^2 b)}{b+\delta^2 b(s+b)}\right) -
  \frac{1}{\delta^2 }\ln\left(1+\delta^2\frac{s}{1+\delta^2 b}\right)\right]} \\
   \mathcal{Z}_\text{excl} &=\sqrt{2\left[s-b\ln\left(\frac{b+s+x}{2b}\right)
  - \frac{1}{\delta^2 }\ln\left(\frac{b-s+x}{2b}\right)\right] -
  \left(b+s-x\right)\left(1+\frac{1}{\delta^2 b}\right)},
 \ee
with
 \be
 x=\sqrt{(s+b)^2- 4 \delta^2 s b^2/(1+\delta^2 b)}.
  \ee
Here, $s$ and $b$ are the expected events of the signal and total SM background after all cuts,  respectively,  and $\delta$ is the percentage systematic  error. To illustrate the effect of systematic uncertainty on the significance, we select three cases: no systematics ($\delta$=0) and two typical systematic
uncertainties~($\delta$=5\% and $\delta$=10\%).
In the limit of
$\delta \to$ 0,  these expressions  can be simplified as
\be
 \mathcal{Z}_\text{disc} &= \sqrt{2[(s+b)\ln(1+s/b)-s]}, \\
 \mathcal{Z}_\text{excl} &= \sqrt{2[s-b\ln(1+s/b)]}.
\ee

\begin{figure}[htb]
\begin{center}
\vspace{-0.5cm}
\centerline{\epsfxsize=8cm \epsffile{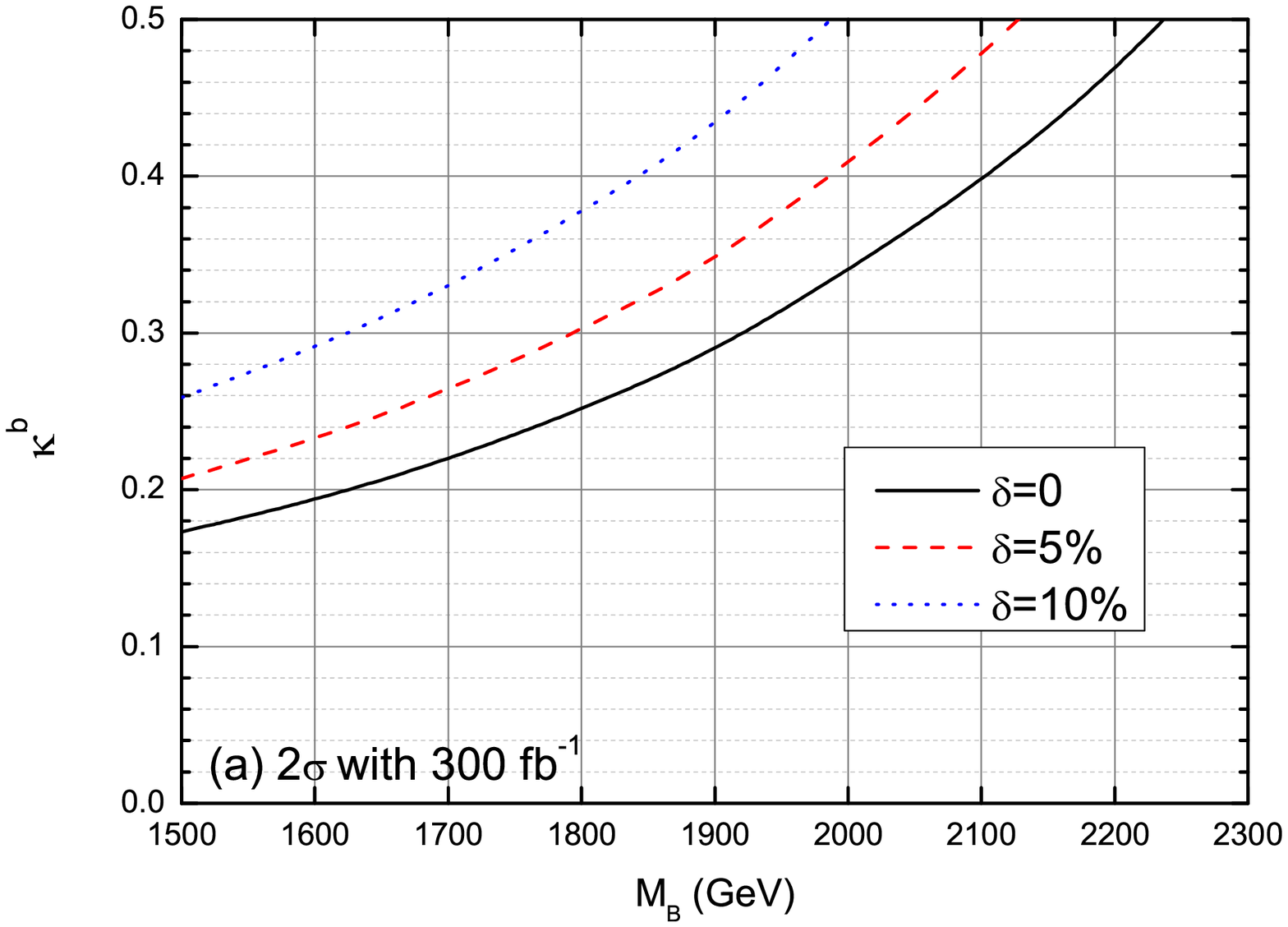}\epsfxsize=8cm \epsffile{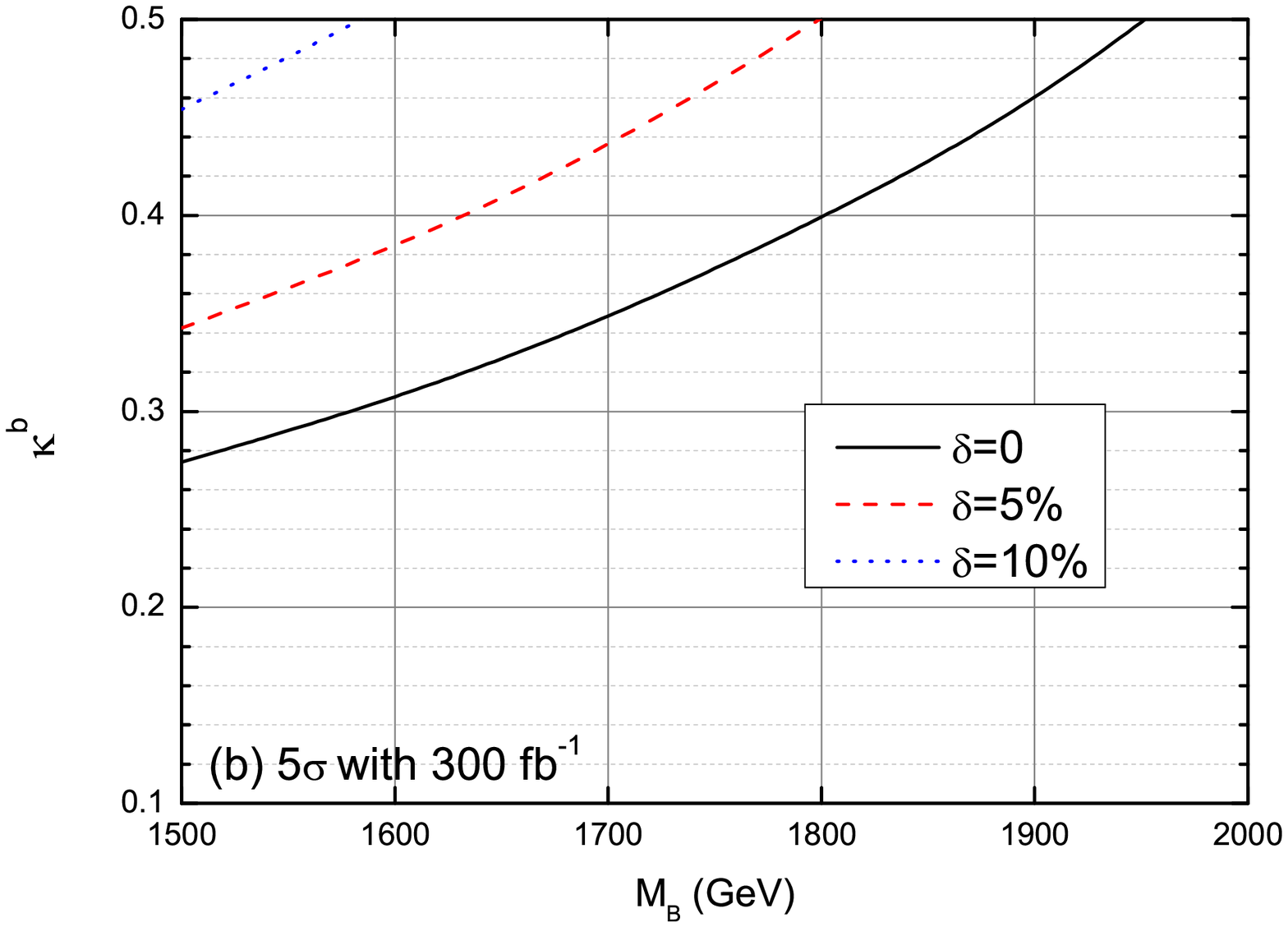}}
\caption{The exclusion limit (at $2\sigma$)  and discovery prospects (at $5\sigma$) contour plots for the signal in $\kappa^{b}-M_{B}$ planes at 14 TeV LHC with integral luminosity of 300 fb$^{-1}$.}
\label{ss1}
\end{center}
\end{figure}
\begin{figure}[htb]
\begin{center}
\vspace{-0.5cm}
\centerline{\epsfxsize=8cm \epsffile{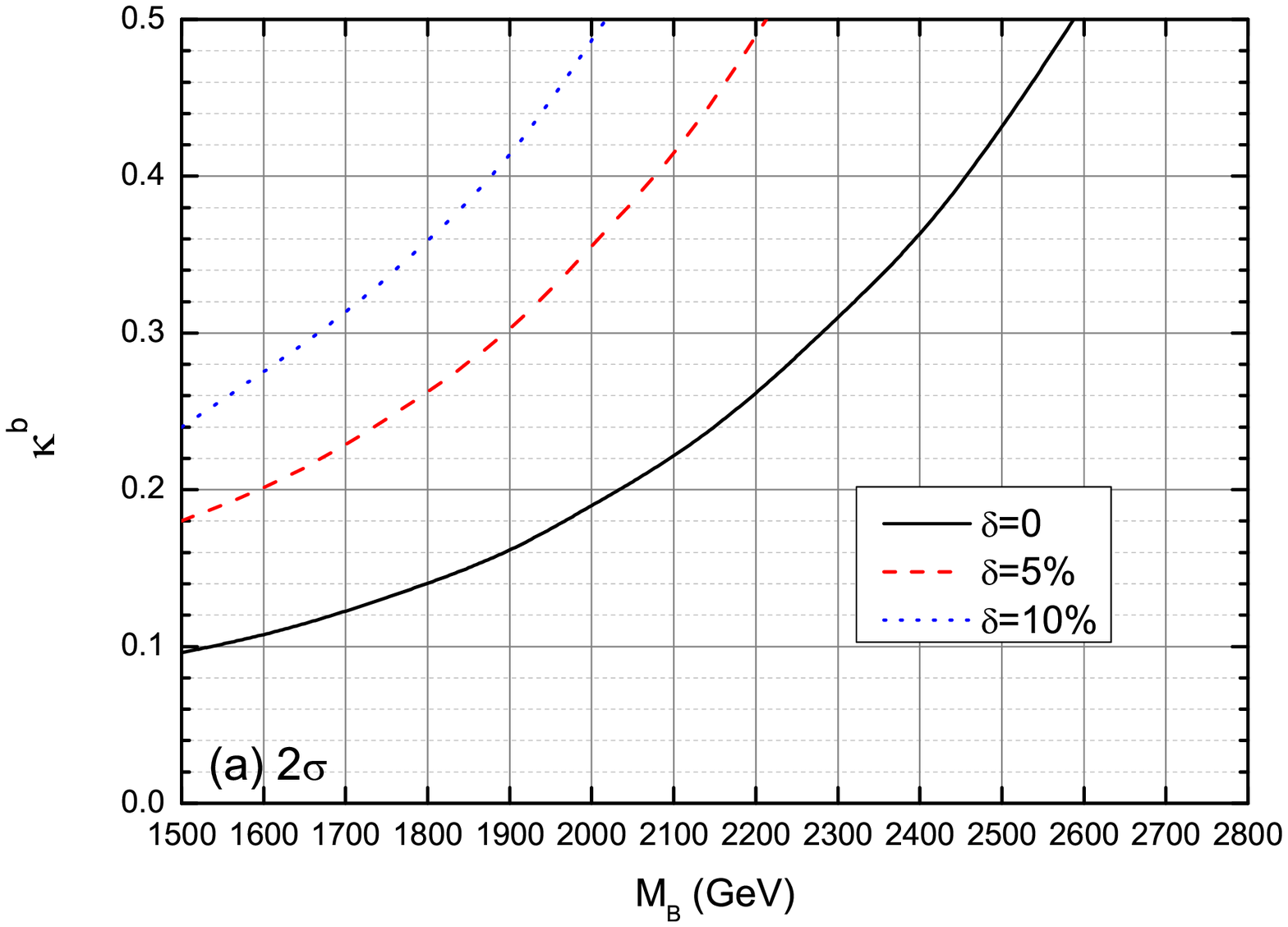}\epsfxsize=8cm \epsffile{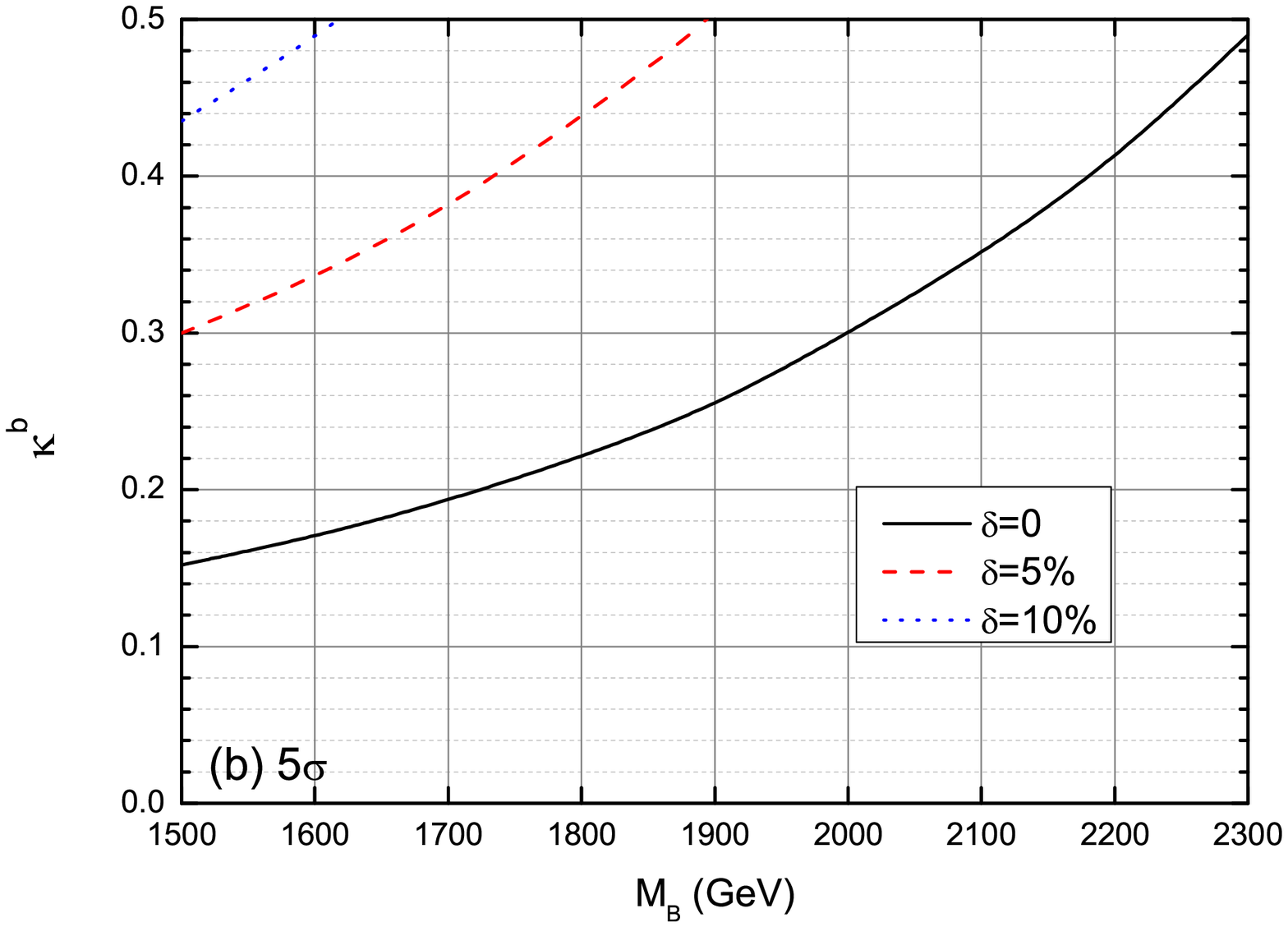}}
\caption{Same as Fig.~6, but for the integral luminosity of 3000 fb$^{-1}$.}
\label{ss2}
\end{center}
\end{figure}

In Fig.~\ref{ss1} and Fig.~\ref{ss2}, the $2\sigma$ and $5\sigma$ lines are plotted as a function of
$\kappa^{b}$ and the VLQ-$B$ mass $M_B$ with the aforementioned three systematic error cases of  $\delta$=0, 5\% and 10\%, for two fixed values of integrated
luminosity 300 and 3000 fb$^{-1}$, respectively. We observe that  our signals are rather robust against the systematic uncertainties
on the background determination.
 For $\kappa_{b}=0.5$ and $\delta=0$, VLQ-$B$ can be probed at the $5\sigma$ level with mass approximately 1950~(2300)~GeV with the integral luminosity of 300~(3000)~fb$^{-1}$, while the $2\sigma$
   exclusion limits are approximately 2230~(2600)~GeV with the integral luminosity of
    300~(3000)~fb$^{-1}$.
        When a realistic 5\% systematic uncertainty is considered, the $5\sigma$  sensitivity decreases to 1800~(1900)~GeV with the integral luminosity of 300~(3000)~fb$^{-1}$. Meanwhile, the $2\sigma$
   exclusion limits decreases to  2100~(2200)~GeV with the integral luminosity of
    300~(3000)~fb$^{-1}$.  Altogether, the higher systematic uncertainty of background can decrease the discovery capability and the excluded region.

We can now
perform a comparison with other complementary studies for searches at the LHC run II
involving a resonant VLQ-$B$. In Ref.~\cite{Gong:2019zws}, the authors  use the same assumption on the parameters and design a dedicated search
strategy for the  $pp\to B\to bZ$ process via the $B\to Zb$ decay mode with $Z\to \ell^{+}\ell^{-}$ at $\sqrt{s }= 14$ TeV with an integrated luminosity of 300~fb$^{-1}$, and the numerical results show that for $\kappa_{b}=0.5$, VLQ-$B$ can be probed at the $5\sigma$ level with a mass approximately 1620~GeV and an integral luminosity of 300  fb$^{-1}$.  Using the model that includes a topless vectorlike doublet $(B, Y)$, the author of \cite{Choudhury:2021nib} investigates the single production of a VLQ-$B$~\footnote{Note that this case using different assumptions on the parameters, the decay $B\to Wt$ is highly suppressed for the small mixing parameter $s_{L}\ll s_{R}$.}, with the $B
  \to b + Z/H$ subsequent decay channel and the fully hadronic final states using a jet substructure at the 13~TeV LHC with an integrated luminosity of 300~fb$^{-1}$; a modest value of the chromomagnetic transition moments ($\kappa =0.5$) enables for the exclusion of $M\lesssim 1.8 \, (2.2)
  ~\rm TeV$ in the $Z$ and $H$ channels respectively.
Therefore, our analysis is competitive with the results of these
existing literature, and represents a complementary candidate to search
for a possible VLQ-$B$.

Note that to avoid having to reject the overwhelming QCD multi-jet background, we do not consider other types of final events, such as the lepton+jets channel~(where only one $W$ boson decays leptonically and the other one decays hadronically) and all hadronic channel~(where both of $W$ bosons decay hadronically). Because these channels are dependent on different tagging algorithms for the identification of boosted, hadronically decaying, heavy particles~\cite{Kaplan:2008ie,Lapsien:2016zor,CMS:2020poo}, i.e., to reconstruct the top quark, a new tagging technique such as the heavy object tagger with variable R~(HOTVR)~\cite{Lapsien:2016zor} algorithm,  can be used to identify jets from the collimated top quark decay, which
is beyond the scope of this paper.

\section{Conclusion}
Based on a simplified model to describe the interactions $Bgb$,  we have investigated the potential for the 14~TeV LHC to discover the vectorlike bottom quark partner produced via its chromomagnetic moment interaction, $pp\to bg\to B$,  with the subsequent decay channel $B\to tW$ and leptonic decay mode for the top quark and $W$ boson. This production mechanism complements the traditional searches which have relied on pair-production of vectorlike quark states, or single production of these states through electroweak interactions.
Our numerical results show that, with a benchmark coupling parameter $\kappa^b=0.5$,  the VLQ-$B$ can be probed at the $5\sigma$ level with a mass approximately 1950~(2300)~GeV and an integral luminosity of 300~(3000)~fb$^{-1}$.
 Meanwhile, the 2$\sigma$ excluded region is of approximately 2230~(2600)~GeV with an integral luminosity of 300~(3000)~fb$^{-1}$. We expect that our analysis can represent a complementary candidate to search
for such singlet VLQ-$B$ quark at the upgraded LHC.

\begin{acknowledgments}
The work is supported by the Foundation of the Key Research Projects in Universities of Henan:22A140019, the Natural Science Foundation of Henan Province:222300420443, and the Project of Innovation and Entrepreneurship Training for College Students in Henan Province (S202110478024).
\end{acknowledgments}

\end{document}